
\input amstex
\documentstyle{amsppt}
\font\special=cmr6
\define\DET{\text{det}}

\define\la#1{\lambda_{#1}}
\define\laaa#1{\lambda_1 ,\cdots , \lambda_{#1}}
\define\gaaa#1{\gamma_1,\cdots,\gamma_{#1}}
\define\teet#1#2{\theta [\eta _{#1}] (#2)}
\define\tede#1{\theta [\delta](#1)}
\define\taaa #1{\tau _1,\cdots, \tau _{#1}}
\define\eppp #1{\epsilon _1,\cdots, \epsilon _{#1}}
\def\I{{\Bbb I}}
\def\R{{\Bbb R}}
\def\C{{\Bbb C}}
\def\Z{{\Bbb Z}}
\def\Q{{\Bbb Q}}
\def\?{\wedge}
\def\LOG{\text{log}}
\def\O{{\Cal O}}
\def\N{{\frak N}}
\def\M{{\frak M}}
\NoRunningHeads
\TagsAsMath
\TagsOnRight
\NoBlackBoxes
\topmatter
\title
What are we quantizing in integrable field theory?
\endtitle
\author
Feodor A. Smirnov\\
{\special Research Institute for Mathematical Sciences,
Kyoto  University,}\\{\special Sakyo-ku 606, Kyoto, JAPAN}\\
{\special and}\\
{\special Steklov Mathematical Institute, St. Petersburg 191011,  RUSSIA}
\endauthor
\dedicatory
Dedicated to my teacher Professor L.D. Faddeev on his forthcoming
sixtieth birthday.
\enddedicatory
\abstract
We continue study of the connection of classical limit
of integrable asymptotically free field theory to the
finite-gap solutions of classical integrable equations.
In the limit  the momenta of
particles turn into the moduli of Riemann surfaces while
their isotopic structure is related to the period lattices.
In this paper we explain that the classical limit of the
local operators can be understood as a measure induced on
the phase space by embedding into the projective space
of "classical fields".
\endabstract
\endtopmatter

\head Introduction \endhead
The present paper develops the ideas of the paper [1]. Let us remind
the basic points of [1]. One of the most important achievements in
the understanding of the structure of integrable models in the last two years
is in the realization of the fact that the bootstrap equation [2] for the
form factors for certain models can be considered as deformed Knizhnik-
Zamolodchikov equations [3,4]. Thus the very possibility of the exact solution
can be considered as just a result of the possessing infinite dimensional
quantum symmetry.

Certainly, we should try to develop deeper understanding
of this situation. In particular it is important to realize what happens
in the classical limit when the infinite-dimensional quantum symmetry
turns into classical one. Formally the limit of the kind is possible
for asymptotically free theories, it corresponds to the limit
from L\"uscher's nonlocal charges [5,6] into the dressing symmetries of
classical integrable models [7]. As it is explained in [1] such a limit
destroys the structure of space-time of the quantum theory.
Effectively at every point of the space-time we get a hierarchy
of classical finite-dimensional systems.
These finite-dimensional systems correspond to $n$-particle subspaces.
Different points are isolated
from each other before the quantization, only after the quantization
is performed they
start to interact through the exchange of particles. It means that
the relativistic space-time appears as a result of quantization
which sounds as an amusing way to solve the main problem of
quantum field theory i.e. to combine relativistic and quantum physics.

The classical analogues of momenta of particles in that approach
are the moduli of
classical solutions of the finite-dimensional systems in question
which happen to be different stationary (finite-gap) solutions
related to Riemann surfaces [8] of
classical integrable equations,
and the isotopic structure is related to the period lattices.
More complicated question is what is
a classical analog of local operators of the quantum model (more
exactly of generating operator of the local operators, see [1] for
details). The answer proposed in [1] is not quite
satisfactory, so we have to return to the question. In the present
paper we shall argue that the local operators in classics
provide a measure on classic trajectories (Jacobi varieties) induced
by their embedding into projective spaces of classical
fields.

\head 1. Theta Formula for the Solutions of KZ \endhead

As it has been shown in [4] the form factors of the generating
function of local fields (from which energy-momentum tensor and
currents can be obtained) for $SU(2)$ Thirring
(chiral Gross-Neveu) model can be
considered as invariant solutions of
Yangian deformation of KZ equations for spin 1/2
vertex operators on level zero. Similar fact is true for
other relativistic models with rich quantum symmetries [2] as well as for
lattice models [9]. The appearance of zero central extension is absolutely
universal in the context and deserves special attention.

Let us mention two basic points. First, the arguments of the equations are
the rapidities of particles. Second, the inner automorphism
of the quantum algebra (antipode square) which is the analog of $L_{-1}$
for the usual equations is identified with the complete rotation of the
space-time around the point where the local operator lives. That means
that we are doing not the deformation of CFT, but study completely
different application of KZ equations. Also that means that in the classical
limit (Yangian double $\to\ \widehat{sl}(2)$) the structure of the space-time
is lost (antipode square which corresponds to topological operation
turns into infinitesimal generator $L_{-1}$ ). These problems are discussed
in details in [1]. The main conclusion is that
in the limit the theory splits
into finite-dimensional systems which are finite-gap solutions of
KdV i.e. are related to different hyper-elliptic surfaces.

Let us remind the most important formula from [1] which presents the
solutions of level zero KZ equations (into which the form factor
equations turn in the classical limit) in terms of Riemann $\theta$-functions.
The KZ equations in our case look as

$$\bigl( \frac{d}{d\la i}+ \sum_{i\neq j}r_{i,j}
(\la i - \la j)\bigr)f(\laaa {2g+2})=0 $$
where $r$ is the classical r-matrix:
$$r_{i,j}
(\la i - \la j)=\frac {\sigma^a_i\otimes \sigma^a_j}
 {\la i-\la j}$$
The vector function $f(\laaa {2g+2})$ belongs to
$({\C} ^2)^{\otimes (2g+2)}$.
We consider real $\laaa {2g+2} $ for they correspond to
the limits of rapidities, also we require $\la 1<\cdots <\la {2g+2}$.
There are many solutions to the equations which are parametrized by
different choices of the sets of $g$ independent contours
($c_1,\cdots,c_g$) on the hyper-elliptic
surface $\Sigma$ defined by the equation
$\tau ^2=\prod (\lambda -\la i)$. Those solutions are of the
main interest for which $c_i\circ c_j=0$ i.e. which can be used
as half-bases of homology. For such solutions the components of
the vector $f_C (\laaa {2n})$ appears to be the special values of
one holomorphic function of $g$ variables (function on the
Jacobi variety of $\Sigma$). This function is expressed in terms of
Riemann $\theta$-function. Namely, consider $\theta$-function
$\theta _C (z)$ constructed respectively to the half-basis
$C=\{c_1,\cdots,c_g\}$, the variable $z\in {\C}^g$. Then one has
$$f_C (\laaa {2g+2})^{\epsilon _1,\cdots,\epsilon _{2g+2}}=F(z)|_
{z=a(\epsilon _1,\cdots,\epsilon _{2g+2})} \tag 1 $$
where $\epsilon _i=\pm$ is $\C ^2$ index,
$$F(z)=D\ \theta _C ^4(z)\DET \bigl[\partial _{z_i}  \partial _{z_j}
\LOG \theta _C (z) \bigr]_{g\times g} \tag 2 $$
The constant $D$ in (2) depends on $\laaa {2g+2}$ but does not depend
on  $\epsilon _1,\cdots,\epsilon _{2g+2}$.
Finally,  $a(\epsilon _1,\cdots,\epsilon _{2g+2})$ is the following
half-period:
$$a(\epsilon _1,\cdots,\epsilon _{2g+2})
= \eta ''(\epsilon _1,\cdots,\epsilon _{2g+2}) +
\Omega _C  \eta '(\epsilon _1,\cdots,\epsilon _{2g+2})=
\sum _{k=0}^g \int\limits _{\la {2k+1}}^{\la {i_k}}\omega _C$$
where  $\Omega _C$ is the period matrix constructed with respect to $C$,
$ \omega _C $ are corresponding normalized first kind differentials,
$\{i_k\}_{k=0}^g$ are ordered numbers for which $\epsilon _{i_k}=+$,
the vectors $\eta ',\eta ''\in {1 \over 2} \Z^g$.

Let us fix the canonical choice of the homology basis. We put the
cuts on the plane between $\la {2i-1}$ and $\la {2i}$. Then the
$a$-cycle $a_i$ starts from the upper bank of $(i+1)$-th cut,
goes to the upper bank of the $i$-th cut, then crosses it
and returns to the starting point by another sheet. The cycle
$b_i$ is taken as the sum of cycles around the $j$-th cuts
for $1\le j \le i $. This choice differs from one used in [1],
but it is more appropriate for the connection with finite-gap
integration: KdV angles vary over the product of $a$-cycles.

The solution to KZ which describes the asymptotics of the
form factor corresponds to $C=\{b_1,\cdots,b_g\}$.
Certainly, the formula (2) can be rewritten in terms of one
canonical $\theta$-function, that defined respectively to
$a$-cycles (corresponding $\Omega$ is pure imaginary).
If $C$ is related to $A$ by
$$\pmatrix {\Cal A}, &{\Cal B}\\ {\Cal C}, & {\Cal D} \endpmatrix \in
Sp(2g,{\Z})$$
then [10]
$$ F_C(z)=\theta ^4 (z)\DET \bigl [({\Cal D}+\Omega {\Cal C})_{i,k}
\partial _{z_k}  \partial _{z_j}
\LOG \theta (z) +2\pi i{\Cal C}_{i,j} \bigr]_{g\times g} $$
In particular the limit of the form factor (which we denote by $f$
without index) corresponding to $b$-cycles is related to the following $F$:
$$ F(z)=\theta ^4 (z)\DET \bigl [\Omega _{i,k}
\partial _{z_k}  \partial _{z_j}
\LOG \theta (z) +2\pi i\delta_{i,j}\bigr]_{g\times g} \tag 3$$
$\delta_{i,j}$ in the last formula is the Kroneker symbol.
It is clear that in order to understand the real meaning of the
classical limit in question we have to understand the meaning of
the function $F(z)$.
We argue that the limit is connected with the finite-gap integration.

Let us make several general remarks. The formulae (1),(2) describe
the classical limit of the form factors for $SU(2)$ Thirring
model in quite special terms. To every multiparticle state a hyper-elliptic
surface is related such that the rapidities of particles are considered as
the position of branching points (moduli) and the isotopic structure
is given by even non-singular half-periods of the Jacobi variety (Jacobian).
One can imagine that this situation is not restricted to the hyper-elliptic
case, and that it might be possible to construct integrable theory
for which the same amount of data (moduli and even nonsingular half-periods)
taken for more general surfaces will describe the space of states,
and the classical limit of matrix elements of local
fields will be given by $F(z)$. The
restrictions of such theory onto
$Z_N$-invariant surfaces will give
$SU(N)$-invariant Thirring models.

Another point concerns $\theta$-functions. Why is it important to
have a description of the classical limit in terms of $\theta$-functions?
The answer to this question becomes clear if we think about the structure
of classical integrable models. The angle-variables should constitute
real torus due to Liouville's theorem. But the mechanism for
the integrability in all interesting cases is Abel transformation.
So, the torus in question allows extension
to a complex torus. Doing any
particular model we have to start with some not too complicated
phase space, and then to embed the Liouville torus into this
phase space. But in practice this construction allows complexification
and the embedding happens to be holomorphic.
The only object which can be used for
the holomorphic embedding of complex torus into a reasonable phase space
is Riemann $\theta$-function. For that reason doing quantization
of integrable models sooner or later we have to come upon the
quantization of $\theta$-function.

\head 2. Connection with finite-gap solutions \endhead

The situation with which the finite-gap theory [8] deals can be summarized
as follows. The infinite-dimensional system (KdV) has finite dimensional
orbits. The motion on the latter is described as the motion
of $g$ real points $P_1,\cdots,P_g$  situated on different
segments $\la {2i}\le P_i\le \la {2i+1}$. In other words the divisor
$P_1,\cdots,P_g$ runs over the product of $a$-cycles on the curve $\Sigma$.
The dynamics is linearized by the Abel transformation which maps
the motion above into the motion over the real
$g$-dimensional subtorus ($J^{\R}=a_1\times\cdots\times a_g$) of
the Jacobian  ($J$) which is complex $g$-dimensional torus.
The dynamical meaning of $J^{\R}$ is clear: it is the torus of angles of
the integrable system. Locally (with respect to actions) the phase
space of integrable system with $2g$ degrees of freedom
looks as $T^g\times {\R}^g$ where $T^g$ is
the torus of angles and the actions vary over ${\R}^g$. In the theory of
KdV the action variables depend on the moduli (positions of the
branching points). Here it is quite undesirable for us to vary the
moduli, there are two possibilities
to avoid doing this.
One way is to consider much bigger phase space, and then to apply
constraints. Another way is to consider the phase space locally.
Namely, consider not the phase space itself
but the product of coordinate space ($J^{\R}$) by the cotangent
space in the direction of angle variables (${\R}^g$ with the
basis $\xi _1,\cdots ,\xi _g$). This manifold is enough to
write differential forms and things like that. Let us mention one
important circumstance. We use the real part of the Jacobi
variety, but it allows a natural complexification. One can try to
use this circumstance in order to introduce certain structures
on $J^{\R}\times {\R}^g$ inducing them from the complexification.
i.e. from $J=J^{\R}\times J^{\I}$, the imaginary part  $J^{\I}
=b_1\times\cdots\times b_g$ .

So the space of the classical trajectories
($\N$) is the same as the collection of real parts of all the Jacobians
of hyper-elliptic surfaces
with real branching points. This space has singularities
of double origin: first, the number of branching points
can become infinite, second, singularities appear when
two branching points coincide. It should be emphasized that
we must not worry about these singularities
on the classical level: the quantization takes care of them,
they become irrelevant in quantum model.

Let us consider the affine model of $\N$. Take the set of all
$2\times 2$ traceless, real, polynomial in additional real
parameter $\lambda$ matrices with the leading coefficient
fixed to be $\sigma ^3$. Factorize this set by the ajoint action
of the matrices $\text{diag}(a,\ a^{-1})$. The set of such
matrices splits into the orbits $\O _{\laaa {2g+2}}$:
every such orbit consists of the matrices with fixed determinant
$$\DET N(\lambda) = \prod \limits _{i=1}^{2g+2}
(\lambda -\la i) $$
The affine model of $\N$ (denoted by $\N _a$) coincides with the joint of
all such matrices with real $\laaa {2g+2}$. $\N _a$ is indeed
a model for $\N$ since the orbit $\O _{\laaa {2g+2}}
$ is parametrized by
the real part of the Jacobian associated to the curve with the
branching points  $\laaa {2g+2}$. Exact description of that
is given later ((4),(5),(6)). From the point of view of finite-gap
integration such matrix describes
the $M$-operators associated with the stationary time.

Now let us map the space $\N _a$ into a bigger space $\M$.
The points of the latter space are
$$\{\laaa {2g+2},\ \psi _1,\cdots ,\psi _{2g+2}\}$$
where $\laaa {2g+2}\in \R,\quad\psi _i$ is vector from $\C ^2$:
$$\psi _i =\pmatrix \alpha _i \\ \beta _i \endpmatrix$$
Take some $N(\la {})\in \N _a$ with the determinant
$ \DET N(\lambda) = \prod
(\lambda -\la i)
$. and consider $N(\la {i})$ ($i=1,\cdots, 2g+2$).
Evidently being a matrix with zero trace and zero
determinant it can be presented as
$$\align &N(\lambda _i)= P' (\la {i})\bigl( \psi _i
\otimes\bar { \psi} _i \bigr)=
\pmatrix \alpha_i\beta_i,&-\alpha_i^2\\
\beta_i ^2, &- \alpha_i\beta_i
\endpmatrix ,\\
&\bar { \psi} _i=\psi _i ^t c,\qquad
c=\pmatrix 0,& -1\\1,&0\endpmatrix\endalign $$
for some $\psi _i$ ( $P' (\la {i})\equiv \prod _k ' (\la i-\la k)$ is
introduced for normalization). This describes
(up to $\psi _i \to \pm\psi _i$) the map
$\N _a \to \M$.

We consider the space $\M$ as the phase space with canonical
Poisson structure given on every finite-dimensional
subspace by
$$\omega =\sum d\alpha _i\wedge d\beta _i$$
the variables $\{\laaa {2g+2}\}$ are just constants
(Poisson commute with everything). Now we want to proceed in
opposite direction: to describe $\N _a$ as join of trajectories
of Hamiltonian systems defined on $\M$. Consider the following
momentum map
$$
\{\laaa {2g+2},\ \psi _1,\cdots ,\psi _{2g+2}\}\to N'(\lambda)=
\sum\limits _{i=1}^{2g+2}{1 \over {\la {}-\la i} } \psi _i\otimes
\bar{\psi _i}  \tag 4
$$
The approach to integrable equations using this momentum map is
presented in [10], see also the recent paper [11].
With the canonical Poisson structure for $\psi _i$ the matrices $N '$
satisfy r-matrix Poisson brackets [12] which provides that their determinants
are in involution. The determinants can be written in the form
$$
\DET (N'(\la {}))=\frac {P(\la {})} { \prod (\la {}-\la i)},\qquad
deg(P)\le 2g
$$
Generally for fixed values of $2g+1$ integrals (fixed $P(\lambda)$)
we have $2g$-dimensional torus of angles which coincides with
the Jacobian of the surface $$\tau ^2=P(\la {})\prod (\la {}-\la i)$$
But these are not generic
orbits we are interested in. Oppositely, let us consider
completely reduced orbits for which $P(\lambda)=1$. Then the only
special points of our matrix are $\laaa {2g+2}$. If we do not
impose this constraint every Jacobian will be counted many times.
Now let us consider
$$N(\la {})=\prod (\la {}-\la i)N'(\la {}) \tag 5 $$
which is a polynomial matrix.
Every completely reduced orbit is organized as
$(J^{\R}\  associated\  to\ the\ curve\ \tau ^2=\prod (\la {}-\la i) )
\ \times
(gauge\ transformations)$. Let us explain this. On every
completely reduced orbit a non-degenerate subset exists
of the matrices whose leading coefficient is not
degenerate as $2\times 2$ matrix (and, hence, stands
before $\lambda ^{g+1}$). This subset is parametrized by $J^{\R}$.
Other elements of the orbit are obtained from this
subset by similarity transformations (gauge transformations)
with polynomial matrix
whose determinant equals 1, for example
$$\pmatrix 1,&Q(\lambda)\\ 0,&1\endpmatrix,\qquad Q(\lambda)
\ \text{is polynomial}$$
So, to map the completely reduced orbit into $\N _a$ one
has to take the non-degenerate subset only (to consider
one representative in every gauge class). This procedure
allows usual Hamiltonian interpretation: in the polynomial
$P(\lambda)$ there are $2g$ coefficients in involution,
on the completely reduced orbit $g$ of them work as Hamiltonians
(govern the motion along $J^{\R}$) the remaining $g$ should be
treated as first kind constraints.

Let us describe explicitly the map from $J^{\R}$ into
$\M$ which is relevant for the construction above:
$$x\to \{\psi _j(x)\}_ {j=1}^{2g+2},\qquad \psi _j(x)=
\frac 1 {\theta (2x)}\pmatrix \teet j {r+2x}\\ \teet j {r-2x}
\endpmatrix \tag 6$$
where $\eta _j$ is the half period defined by the integral
from fixed branching point (say, $\la 1$) to the point $\la j$,
$$r=\int\limits _{\la 1} ^ {\infty ^+} \omega$$
$\infty ^+$  is one of the infinities on the surface, actually, any
other real non Weierstrass point would do with minor changes, the
thing which does matter is that the shift of arguments in two components of (6)
corresponds to a singular divisor
i.e. two different points on the
surface which project onto the same
point on the complex plane. We put $2x$ in the argument of
$\theta$-functions in order that the map is properly defined on the
Jacobian. Comments on the formulae (6) are given in Appendix.

Let us now, as in [1], using the solution of KZ construct the
following homogeneous polynomials on $\M$ :
$$P(\psi)=\bar{\psi}_{1,\epsilon _1}
\otimes\cdots\otimes \bar{\psi}_{2g+2,\epsilon _{2g+2}}
\  f(\laaa {2g+2})^{\epsilon _1,\cdots,\epsilon _{2g+2}} \tag 7
$$
which is just the scalar product of two vectors defined
in  ${\C}^{\otimes (2g+2)}$.
The formula (7) is constructed by analogy with the quantum case:
it is similar to the contribution of $2g+2$ particles with
rapidities $\laaa {2g+2}$ into the generating function of local
operators. The vectors $\psi _i$ play role of creation operators of
two-component particles. The idea of considering this object is
the following. The space of particles in the quantum theory is
"free" and rather huge (similar to classical $\M$). The local operators
cut some pieces from this space. In classics that should correspond
to considering of (7) on the equation of motion i.e. we want to
consider $P(\psi (x))$ with $\psi _i(x)$ given by (6).

So the problem is to calculate $P(\psi (x))$ using the formulae (1),(2),(6).
We expect the following result
$$P(\psi (x))= Const\ \DET \bigl [\Omega _{i,k}
\partial _{x_k}  \partial _{x_j}
\LOG \theta (2x) +2\pi i\delta_{i,j}\bigr] _{g\times g} \tag 8 $$
The last formula is not easy to prove, however there is
strong evidence in favor of it.
First, considering the vectors $f_C$ corresponding to all
possible half-bases $C$ one makes sure that the modular
properties of RHS and LHS of (8) are the same.
Second, it is easy to realize that  $P(\psi (x))$ does provide an interpolation
for the RHS for all even nonsingular half-periods. As it has been said
these half-periods correspond to partitions of $\Lambda=
\{\laaa {2g+2}\}$ into $\Lambda ^+$ and $\Lambda ^-$ such that
$\#\Lambda ^+=\#\Lambda ^-=g+1$. The matrix $N(\la {})$ for such
half-period is given by
$$U\pmatrix 0,&\prod\limits _{\la i \in \Lambda ^+} (\la {}-\la i)\\
\prod\limits _{\la i \in \Lambda ^-} (\la {}-\la i),
&0\endpmatrix U^{-1}\tag 9 $$
for some constant matrix $U$ with $\DET U=1$. So, the values of $\psi _i$
at these points are quite simple, the matrix $U$ can be omitted
when substituting into (7) because  $f(\laaa {2g+2})$ is singlet
(spin zero) vector in the tensor product. It is explained in
the Appendix how to get (9) from (6).

In a sense the
formula (8) inverts the formula (1), it shows that not only $f(\la {1},\cdots,$
$\la {2g+2})$
can be constructed via $F(z)$ but the function from (8) which differs
from $F(2x)$ only by absence of $\theta ^4$ can be constructed via
convolution of $f(\laaa {2g+2})$ with canonical vectors $\psi$.
So, in order to understand the classical limit of the
generating function of local
operators we have to understand the meaning of  $P(\psi (x))$.
Let us think of the explicit formula (8). It looks as a measure
on the phase space.
Recall that locally (in action variables)
we consider  the phase space as angle variable multiplied by infinitesimal
piece of the space of action variables: $J^{\R}\times {\R}^g$.
The cotangent space has the basis $dx_1,\cdots ,dx_g;\xi _1\cdots\xi _g$.
We can write the formula:
$$\DET \bigl [\Omega _{i,k}
\partial _{x_k}  \partial _{x_j}
\LOG \theta (2x) +2\pi i\delta_{i,j}\bigr]_{g\times g}
dx_1 \?\cdots \?dx_g \?\xi _1\?\cdots\?\xi _g=
\?^g\omega $$
where $\omega$ is the following 2-form:
$$\omega =\bigl (\Omega _{i,k}
\partial _{x_k}  \partial _{x_j}
\LOG \theta (2x) +\delta_{i,j}\bigr)dx_i \?\xi _j  \tag {10} $$
So, $P(\psi (x))$ is the maximal degree of the 2-form
$\omega$. We have to understand the meaning of $\omega$. It looks
quite symmetrical, for that reason it is natural
to think of $\omega$ as of the
form induced on the real space
by some K\"ahler structure [13]
on a complexification. That is what we shall explain
in the next section.

\head 3. Tau-function and Lefschetz embedding \endhead

Let us start this section with one formula from the paper [14]:
$$\tau (x+y)\tau (x-y)=\sum F_i(x)G_i(y) \tag {11}$$
where $\tau$ is KdV $\tau$-function, $x=\{x_1,x_2,\cdots\}$ is
an infinite set of times, the functions $G_i$ in the RHS are taken
in "minimal" way (will be clarified soon). In the approach of [14]
$\tau$ is associated with the level 1 representation of $\widehat {sl}(2)$
denoted by $V(\Lambda)$. The LHS of (11) can be thought about as the
tensor product of two such representations. Due to the complete reducibility
the linear hull of LHS for different $y$ is level 2 representation
($V(2\Lambda)$). This linear hull is kept in mind when we talk about
minimality of the set of $G_i$. For  the minimal
choice of $G_i$ the functions $F_i(x)$ constitute the basis
of the representation $V(2\Lambda)$. The latter is realized in
the Fock space
associated to one massless Bose field (Heisenberg algebra)
and one massless Majorana fermion (Clifford algebra) which is a special
case of general parafermionic picture [15].
It is instructive to consider in this situation the
Virasoro central charges which, roughly speaking, count
the number of states in different modules. For the tensor product of
two representations $V(\Lambda)$ the central charge equals $2$,
For the representation $V(2\Lambda)$ the central charge is
${3 \over 2}$ ( $1$ for boson and ${1 \over 2}$ for fermion).
So, for the orthogonal complement of $V(2\Lambda)$ in the tensor
product of two $V(2\Lambda)$ the central charge is ${1 \over 2}$, this
subspace gives rise to all the Hirota equations.

Let us return to the basis in $V(2\Lambda)$ given by $F _i(x)$.
The dependence
on $x=\{x_1,x_2,\cdots\}$
is understood due to
bosonic structure while the index $i$ corresponds to
decomposition in fermions. On the other hand the index $i$ counts all
the nonzero Hirota derivatives of the $\tau$ function. The space of these
derivatives can be considered as the space of different  KdV fields,
but not exactly, due to
Sato we know that
more adequate
understanding of
the space of different  KdV fields is the projective space associated to
the space of Hirota derivatives. For example consider the KdV field $u$ itself.
The corresponding second Hirota derivative (the coefficient before $(y_1)^2$
in Taylor decomposition of (11)) and the  function $\tau ^2(x)$ both are
coordinates in our space. So, $u$ being equal to this second Hirota
derivative
divided by $\tau^2$  is the projective coordinate.
Hence, we think of the map $x \to \{F_i(x)\}$ as of a mapping
of the infinite dimensional abelian group into the projective
space constructed from the fermionic Fock space.

Let us consider the finite dimensional version of this
construction corresponding
to the finite-gap solutions. For them infinite-dimensional
group of $x$ reduces to finite-dimensional one ($J^{\R}$), and
the $\tau$-function is $\theta$-function (usually it is multiplied
by some exponent of quadratic form, but we omit this multiplier
which anyway would disappear from our final formulae).
Not only the group of times but also the space of KdV fields
reduces to finite-dimensional in this limit because for the
decomposition (11) we can use the classical formula:
$$\theta (x+y|\Omega) \theta (x-y|\Omega) =\sum\limits _{a\in {1 \over
2}{\Z}^g}
\theta \left[ {a \atop 0} \right] (2x|2\Omega)
\theta \left[{a \atop 0}\right] (2y|2\Omega)  \tag {12} $$
So, the finite-dimensional version of the space of
KdV fields is the projective space associated to the
$2^g$ dimensional space which evidently can be
related to Clifford algebra with $g$ generators.
The group $J^{\R}$ is mapped into the space via
$$x\to\theta \left[{a \atop 0}\right] (2x|2\Omega) $$
This mapping allows natural complexification which is the mapping
of whole complex torus $J$ into the complex projective space
${\Bbb CP}^{2^g-1}$:
$$z\to w_a(z)=
\theta \left[{a \atop 0}\right] (2z|2\Omega) $$
This is a classical mapping considered by Lefschetz [see 10].

The complex projective space is K\"ahler manifold. It
allows hermitian, riemannian, symplectic structures
which are related in different ways. In particular the
symplectic form is given by
$$\omega=\partial _{w_b}\partial _{\bar{w}_c}\text{log}
\bigl(\sum _a   |w_a|^2\bigr) dw_b \? d \bar{w}_c $$
This form induces symplectic form on $J$:
$$\omega=\partial _{z_i}\partial _{\bar{z}_j}\text{log}
\left(\sum _a
\left|\theta \left[ {a \atop 0}\right] (2z|2\Omega)\right|^2\right)
dz_i \? d \bar{z}_j
$$
The last formula can be simplified using one more time the
addition theorem (12):
$$\sum _a
\left|\theta \left[ {a \atop 0}\right] (2z|2\Omega)\right|^2 =
\sum _a
\theta \left[ {a \atop 0}\right] (2z|2\Omega)
\theta \left[ {a \atop 0}\right] (2\bar{z}|2\Omega)=
\theta (2x|\Omega) \theta (2iy|\Omega) \tag {13} $$
where $x=\frac 1 2 (z+\bar{z}),\  y=\frac 1 {2i} (z-\bar{z})$.
Let us emphasize that the fact that $\Omega$ is pure
imaginary (related to $\laaa {2g+2} \in \R$ and proper choice
of homology basis) is important in this calculation.
Using (13) the symplectic form  on $J$ can be rewritten as
$$\omega=\bigl (\partial _{x_i}\partial _{x_j}\text{log}
\theta (2x|\Omega) +
\partial _{y_i}\partial _{y_j}\text{log}\theta (2y|\Omega)
\bigr) dx_i \? dy_j  \tag {14} $$
The last formula has much in common with (10) but it belongs to
$\?^2T^*(J)$
for the complex torus $J$ while the form (10) belongs to
$T^*(J^{\R})\times {\R}^g$. So, to obtain the form (10) from (14)
one has to map $T^*(J^{{\Bbb I}})$ into ${\R}^g$. This is done
by the period map which relates to any 1-form $\omega^1 \in T^*(J^{{\Bbb I}})$
the following vector from ${\R}^g$ with basis $\{\xi _i\}$:
$$\omega ^1\to \sum _i \bigl(\int\limits _0^{\Omega e_i }
\omega ^1      \bigr) \xi _i $$
where $\Omega$ is the period matrix, $\{e_i\}$ are basic vector of the
lattice  ${\Z}^g$. Using the properties ot $\theta$-function one shows
that under this mapping the form (14) turns into
$$\omega =\bigl (\Omega _{i,k}
\partial _{x_k}  \partial _{x_j}
\LOG \theta (2x) +2\pi i\delta_{i,j}\bigr)dx_i \?\xi _j$$
which coincides with (10).
Thus we have shown that the form (10) is induced on the
local (with respect to actions) phase space by period mapping from
the complex Jacobian, on which the symplectic form
is induced by the canonical mapping into the complex projective
space.

Returning to the general formula (11) we can say the following.
It looks as if our real goal was to introduce a measure on
the embedding of the infinite-dimensional group of times into
the projective space related to
the fermionic part of
$V(2\Lambda)$
(space of classical fields). But we can not
do it directly, so, we split the infinite-dimensional orbit
into finite-dimensional ones and work with them. This is the
very idea of Fock space. It is remarkable that in this way we
are able to perform the exact quantization. It is also important
that after the quantization the contributions from different Jacobians
are connected through the residue equation for form factors [2].

{\bf Acknolegement.}
I would like to thank E. Date,  E. Frenkel,
T. Miwa, \linebreak
A. Reiman, M. Semenov-Tian-Shansky, E. Sklyanin
for useful discussions. Special thanks are due to L. Faddeev for
drawing my attention to K\"ahler geometry.

\head Appendix. \endhead

Here we provide useful information about $\theta$-functions
which can be found in the books [10,16].

1. {\it Definition of theta-function. }
$$\align & \theta (z|\Omega)=\sum_{m\in Z^g}\text{exp}
\{\pi i\ m^t\Omega  m+2\pi i z^tm\}\\ &\teet { }{z|\Omega}=
\text{exp}\bigl\{\pi i\eta^{\prime\prime\ t} \Omega\eta'+
2\pi i(z+\eta'')^t\eta'\bigr\}\theta(z+\eta''+\Omega\eta'|\Omega)
\endalign $$
where $z \in \C^g,\ \eta={\eta' \atop \eta''},
\ \eta',\eta''\in \Q ^g$.

2. {\it Riemann theorem for theta-function on hyper-elliptic surface.}
$$ \theta (\int _{\lambda _1}^{P}-\int _{\lambda _3}^{Q_1} -\cdots
\int _{\lambda _{2g+1}}^{Q_g}) $$
is either identically zero or has simple zeros only at the points
$P=Q_1, \cdots Q_g$ , we use Fay's notations [16]:
$\int =\int \omega $.

3. {\it Divisor of meromorphic function.}
$$\sum \int\limits _{P_i}^{Q_i}  =0$$
if and only if there is a meromorphic function with simple zeros at
$\{P_i\}$ and simple poles at $\{Q_i\}$.

4. {\it Two formulae. }
{}From Riemann theorem and the property of the divisor of meromorphic function
one gets
$$ \align &
\frac {\tede {\int _ {\lambda _1} ^P -\int _ {g\lambda _1} ^{\Lambda _S}}
\tede {\int _ {\lambda _1} ^P +\int _ {g\lambda _1} ^{\Lambda _S}} }
 {\tede {\int _ {\lambda _1} ^P -gr}
\tede {\int _ {\lambda _1} ^P +gr} }=C(S)\prod _{i\in S} (x-\lambda _i)\qquad
 \#S=g\ ;\\&
\frac {\tede
{\int _ {\lambda _1} ^P -\int _ {g\lambda _1} ^{\Lambda _S}+r}
\tede {\int _ {\lambda _1} ^P +\int _ {g\lambda _1} ^{\Lambda _S}-r} }
 {\tede {\int _ {\lambda _1} ^P -gr} \tede {\int _ {\lambda _1} ^P +gr} }=
C(S)(\prod _{i\in S} (x-\lambda _i)-\prod _{i\in \bar{S}}
 (x-\lambda _i)) \\
&\#S=g+1 \endalign $$
where $x$ is the projection of $P$ onto the complex plane,
$\Lambda _S$ is a subset of the set of branching points, $\delta$
is Riemann constant:
$$\delta '' +\Omega \delta ' =\sum \limits _{i=1}^{g+1}\int\limits
_{\la 1}^ {\la {2i-1}}  $$
finally
$$r=\int _{\lambda _1} ^{\infty ^+}=-\int _{\lambda _1} ^{\infty ^-}$$

5. {\it Theta constants.}
The formulae above allow to calculate certain special values of
$\theta$-function. First type of them is given by
$$\teet T 0 =C_1 \prod _{i>j \in T} (\lambda _i-\lambda _j)^{\frac 1 4}
\prod _{i>j \in \bar{T}} (\lambda _i-\lambda _j)^{\frac 1 4}
,\ \#T=g+1,\ \bar{T}=B\backslash T$$
where $T\in \{1,2,\cdots 2g+2\}$,
$$\eta _T '' +\Omega\eta _T '=\int\limits _{\Lambda _U}^{\Lambda _T},$$
the subset $U=\{1,3,5,\cdots\}$ corresponds to Riemann constants. The positive
constant $C_1$ depends on $\{\la i \}$ but does not depend on
the partition $T$, the exact value of $C_1$  is given by Tomae
formula. Another type of $\theta$-constants of interest is given by
$$\teet S r = C_2
 \prod _{i>j\in S} (\lambda _j -\lambda _i)^{\frac 1 4}
\  \prod _{i>j\in \bar {S}} (\lambda _j -\lambda _i)^{\frac 1 4} $$
where $\#S=g$, the characteristic is given by
$$\eta _S '' +\Omega\eta _S '=\int\limits _{\Lambda _{U\backslash
\{1\}}}^{\Lambda_S},$$
$C_1$ is a positive constant.
The second relation allows to prove that the matrix $N(\lambda, x)$
defined by (4),(5),(6) takes at even non-singular half-periods
values (9). The first formula is used when substituting corresponding $\psi _i$
into (7).

6.{\it Frobenius formulae.}

There is the following nice addition formula on
hyper-elliptic surfaces:
$$\sum _{j=1}^{2g+2} (-1)^j
 \theta [\zeta _1+\eta_ j](x_1)
\theta [-\zeta _1+\eta_ j](x_2)
 \theta [\zeta _2+\eta_ j](x_3)
 \theta [-\zeta _2+\eta_ j](x_4)=0$$
for arbitrary characteristics $\zeta _1 ,\zeta _2$,
$x_1+x_2+x_3+x_4=0$, the characteristic $\eta_ j$ is associated to
$j$-th branching point:.
$$\eta _j '' +\Omega\eta _j '=\int\limits _{\lambda _1}^{\lambda _j}$$

Using these formula for $x_1=x_2=0,x_3=-x_4=2x $ and for
$x_1=x_2=-r, x_3=r-2x,x_4=r+2x $ and for proper half-periods
$\zeta _1,\zeta _2 $ one gets after some calculations using the
theta constants above:
$$\align
&\sum\limits _{j=1}^{2g+2}
\lambda_j ^p
\alpha _j^2(x)=0 \ ,\ p=0,\cdots,g
\\&\sum\limits _{j=1}^{2g+2}
\lambda_j ^p
\beta _j^2(x)=0\ ,\ p=0,\cdots,g
\\&\sum\limits _{j=1}^{2g+2}
\lambda_j ^p
\alpha _j (x)\beta  _j (x)=0\ ,\ p=0,\cdots, g-1
\\&\sum\limits _{j=1}^{2g+2}
(2\lambda_j ^{g+1}-(\sum\lambda_i) \lambda_j ^g)
\alpha _j (x)\beta _j (x)=0
\endalign $$
which is needed for proof that the formula (6) gives
parametrization of $ \N _a$.

\end